\documentclass[a4paper,11pt]{article}
\usepackage{pos}
\usepackage{natbib}
\usepackage{caption}
\usepackage{subcaption}

\title{On Discrete Goldstone Bosons}

\author*[a]{Víctor Enguita-Vileta}
\author[a]{Belén Gavela}
\author[b]{Rachel Houtz}
\author[c, d]{Pablo Quílez}

\affiliation[a]{Departamento de Fisica Teorica, Universidad Autonoma de Madrid and IFT-UAM/CSIC,\\ Cantoblanco, 28049, Madrid, Spain}

\affiliation[b]{Institute for Particle Physics Phenomenology, Durham University,
Durham DH1 3LE, U.K.}
\affiliation[c]{Department of Physics, University of California, San Diego,\\
9500 Gilman Drive, La Jolla, CA 92093, USA}
\affiliation[d]{Deutsches Elektronen-Synchrotron DESY
Notkestr. 85, 22607 Hamburg, Germany}

\emailAdd{victor.enguita@uam.es}
\emailAdd{belen.gavela@uam.es}
\emailAdd{rachel.houtz@durham.ac.uk}
\emailAdd{pquilezlasanta@ucsd.edu}

\abstract{Exact discrete symmetries, if non-linearly realized, can reduce the ultraviolet sensitivity of a given theory.  The  scalars stemming from spontaneous symmetry breaking are massive without breaking the discrete symmetry, and those masses are protected from divergent quadratic corrections. This is in contrast to   non-linearly realized continuous symmetries. The  symmetry-protected masses  and potentials of  those {\it discrete Goldstone bosons} offer promising physics avenues, both theoretically and in view of the blooming experimental search for ALPs. In this text, we develop this theoretical setup for the specific case of a triplet of $A_5$ using invariant theory, showcasing the substantial improvements and compelling phenomenological consequences introduced by the invariance under a discrete symmetry.
}

\FullConference{%
  8th Symposium on Prospects in the Physics of Discrete Symmetries (DISCRETE 2022)\\
  7-11 November, 2022\\
  Baden-Baden, Germany
}

\begin{document}
\maketitle

\section{Introduction}

This text is a summary of the work presented in \cite{Vileta:2022jou}, where a new class of light scalar particles is introduced and analyzed in detail. We label them discrete Goldstone bosons (DGBs). 

DGBs are a subset of pseudo Nambu-Goldstone bosons (PNGB) whose scalar potential preserves a discrete symmetry exactly. The main consequence of this invariance is a substantially reduced UV sensitivity in DGBs, in relation to the generic case. More specifically, for a generic PNGB,  quantum corrections quadratic in an UV scale $\Lambda$ may introduce large corrections to their masses and couplings. Meanwhile, corrections of that kind are completely absent in the DGB potential. This translates into more stable masses which contain logarithmic contributions with $\Lambda$ at most. 

Such an improved UV (in)sensitivity offers an array of opportunities in model-building. We highlight the case of composite Higgs models \cite{Kaplan:1983fs,Georgi:1984af,Dugan:1984hq} as solutions to the Hierarchy problem. These models generically lack a mechanism to fully suppress the $\Lambda^2$ corrections to the Higgs potential, often introducing other fine tuning problems in their attempt to solve the initial puzzle. 
Interesting applications may exist also in the context of axions, ALPs, dark matter and other theories containing light scalar particles.

Precedents of our work can be found in Refs.\,\cite{Hook:2018jle,Das:2020arz}.
In our paper, we consider new specific models and identify novel phenomenological features of DGBs. The models we present are proof-of-principle, and we leave more realistic implementions to be presented in further works.

\section{Non-linearly realized discrete symmetries}

Consider a discrete group $D$ and a scalar field $\Phi$ in one of its $m$-dimensional irreducible representations, then
$\Phi \equiv\left(\phi_1, \phi_2 ,\ldots,\phi_m\right)$. If $D$ is spontaneously broken, the degrees of freedom in $\Phi$ are tied up by the quadratic restriction
\begin{equation}
    \Phi^T \Phi=\phi_1^2+\phi_2^2+\ldots+\phi_m^2=f^2\,,
    \label{eq:non-linearity-constraint}
\end{equation}
where $f$ is the scale of spontaneous symmetry breaking (SSB). We call this relation the \emph{non-linearity constraint} and argue that, if $D$ is an exact invariance of the scalar potential $V(\Phi)$, then Eq.\,(\ref{eq:non-linearity-constraint}) is enough to forbid quadratic corrections with UV scales to all levels in the theory. In this text we will show how is this the case by analyzing the particular example in which $ D = A_5$ and $\Phi$ is a triplet. We will also look at the further phenomenological consequences granted by $A_5$-invariance.

How are Goldstone bosons to appear if $D$ is a discrete symmetry, rather than a continuous one? We stress the fact that any discrete group $D$ is embedded in a larger continuous one. In our realizations, the continuous group is an approximate symmetry of the theory, which can be regarded as mildly broken into $D$\footnote{In \cite{Vileta:2022jou}, we develop several example UV completions leading to such a setup, where the $D$-preserving contributions  which break the continuous symmetry explicitly have a power-like suppression with some perturbative coefficient, in a way reminiscent of Clockwork\,\cite{Giudice:2016yja}.}. Therefore the Goldstone Theorem applies.

\section{A triplet of \texorpdfstring{$A_5$}{}}

Among the examples examined in \cite{Vileta:2022jou} is that of the triplet of $A_5$, which has shown to include a number of compelling traits accessory to the UV stability already discussed.

\paragraph{Invariants}

As to find the building blocks of the $A_5$-preserving potential, we use elements of finite group invariant theory. A triplet representation has three elementary invariants called the \emph{primary invariants}; they are such that all the remaining invariants can be written as a function of these three. The \emph{Molien generating function} \cite{molien1897invarianten,burnside1911theory}, in this case given by
\begin{equation}
    \mathcal{F}_{A_5}(\mathbf{1}, \mathbf{3} ; \lambda)=\frac{1+\lambda^{15}}{\left(1-\lambda^2\right)\left(1-\lambda^6\right)\left(1-\lambda^{10}\right)}\,,
    \label{eq:A_5:Molien_function}
\end{equation}
is such that the exponents in its denominator equal the order $n$ of the primary invariants: in this case $n = 2\,,6\,,10$. Since $A_4\subset A_5$, a clever way of writing down these invariants is to write them in terms of those of $A_4$. We write
\begin{equation}
\begin{aligned}
& \mathcal{I}_2=\Phi^T \Phi = \phi_1^2 +\phi_2^2 + \phi_3^2\,, \\
& \mathcal{I}_6=22 \mathcal{J}_3^2+\mathcal{J}_2 \mathcal{J}_4-2 \sqrt{5} \mathcal{J}_6\,, \\
& \mathcal{I}_{10}=\mathcal{J}_2^2 \mathcal{J}_4+38 \mathcal{J}_3^2 \mathcal{J}_4-\frac{7}{11} \mathcal{J}_2^3 \mathcal{J}_4-\frac{128}{11 \sqrt{5}} \mathcal{J}_2^2 \mathcal{J}_6+\frac{6}{\sqrt{5}} \mathcal{I}_4 \mathcal{I}_6\,,
\label{eq:A_5:invariants}
\end{aligned}
\end{equation} 
where $\mathcal{J}_2 = \phi_1^2 + \phi_2^2 + \phi_3^2$, $\mathcal{J}_3 = \phi_1\phi_2\phi_3$, $\mathcal{J}_4 = \phi_1^4 + \phi_2^4 + \phi_3^4$ are the primary invariants of $A_4$
and $\mathcal{J}_6$  fulfils  the \emph{syzygy} $4 \mathcal{J}_6^2=2 \mathcal{J}_4^3-5 \mathcal{J}_4^2 \mathcal{J}_2^2+4 \mathcal{J}_4 \mathcal{J}_2^4-36 \mathcal{J}_4 \mathcal{J}_3^2 \mathcal{J}_2-\mathcal{J}_2^6+20 \mathcal{J}_3^2 \mathcal{J}_2^3-108 \mathcal{J}_3^4$. 

From the numerator in Eq.\,(\ref{eq:A_5:Molien_function}), one also infers the existence of an invariant of order $n = 15$, $\mathcal{I}_{15}$, which can be written as a non-polynomial function of the primary ones. All the remaining invariants of $A_5$ can be written as a polynomial in $\mathcal{I}_2$, $\mathcal{I}_{6}$, $\mathcal{I}_{10}$ and $\mathcal{I}_{15}$. 

\begin{figure}
\centering
\includegraphics[width = 0.65\textwidth]{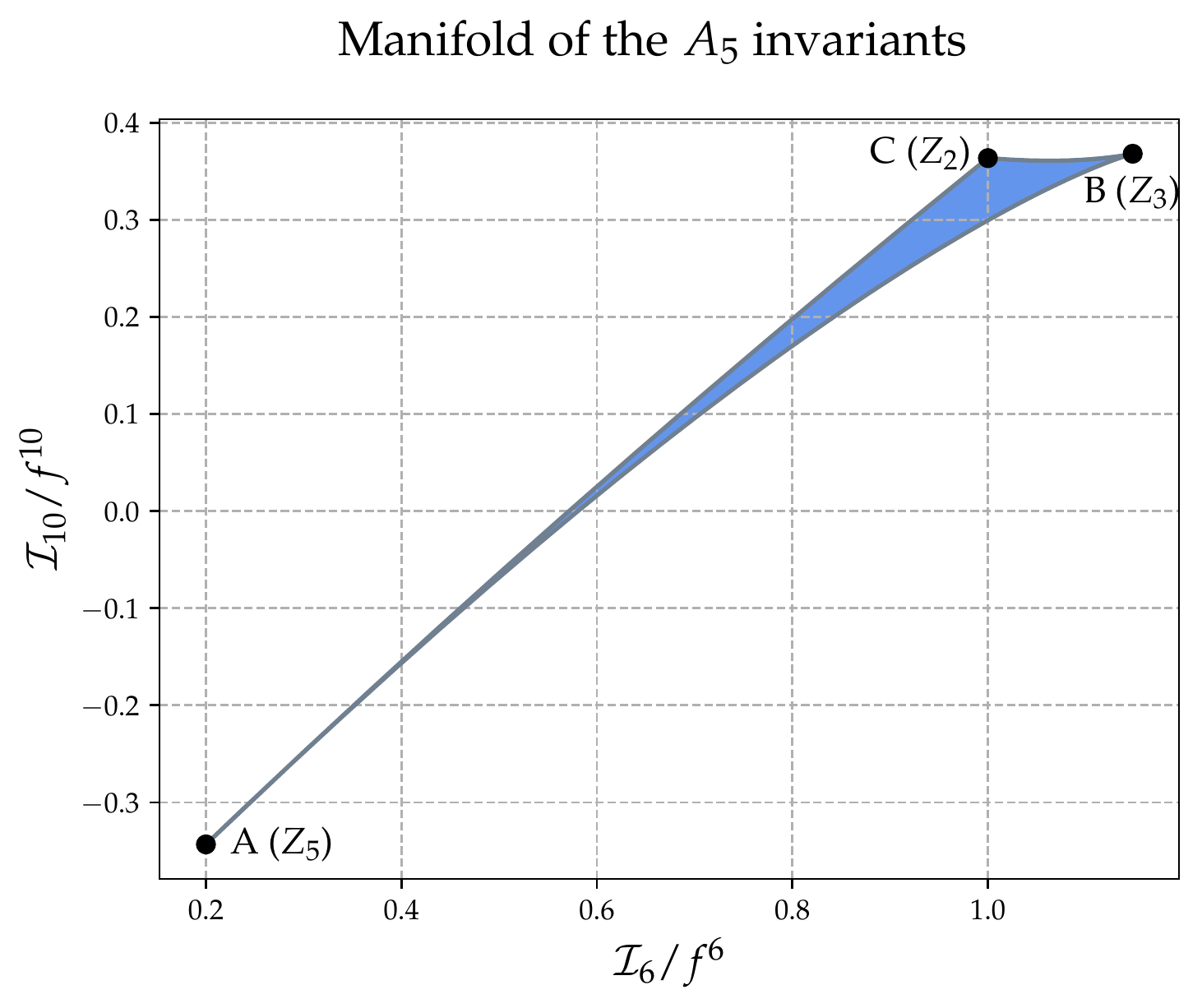}
\caption{Manifold defined by the $\mathcal{I}_{6}(\Phi)$ and $\mathcal{I}_{10}(\Phi)$ invariants of $A_5$, with $\Phi$ a triplet fulfilling $\mathcal{I}_{2}(\Phi)=\Phi^T \Phi=f^2$. A smaller discrete symmetry, denoted in parenthesis, remains explicitly realized after SSB around the MaNa extrema, $A$, $B$ and $C$.}
\label{fig:invariant_manifold}
\end{figure}

\paragraph{EFT}

By Eq.\,(\ref{eq:non-linearity-constraint}), after spontaneous symmetry breaking
$\mathcal{I}_{2} = \phi_1^2 + \phi_2^2 + \phi_3^2 = f^2$.
That is, \emph{the invariant of order $n = 2$ is rendered non-dynamical}. Since $\mathcal{I}_{2}$ is the only invariant of order $n = 2$ allowed by the discrete symmetry, $A_5$ in this case, there are no contributions to to the potential at that order. In consequence, and simply by dimensional arguments, $\Lambda^2$ \emph{contributions must also be absent}. In fact, there will be no renormalizable contributions to this potential, which starts at order $n = 6$.

The complete $A_5$-preserving potential for a triplet $\Phi$ can be written in terms of of $\mathcal{I}_6$ and $\mathcal{I}_{10}$ only,
\begin{equation}
    V_{\mathrm{DGB}}=f^2 \Lambda^2 \sum_{a, b,c}^{\infty}c_{abc}\left(\frac{\mathcal{I}_6}{f^6}\right)^a\left(\frac{\mathcal{I}_{10}}{f^{10}}\right)^b\left(\frac{\mathcal{I}_{15}}{f^{15}}\right)^c\,, \quad \Lambda \leq 4 \pi f\,,
    \label{eq:A5:potential}
\end{equation}
where as stated previously $\mathcal{I}_{15} = \mathcal{I}_{15}(\mathcal{I}_{6}, \mathcal{I}_{10})$.

\paragraph{Maximally natural extrema}

The potential in Eq.\,(\ref{eq:A5:potential}) contains a wealth of possible extrema. They can be generically calculated by solving
\begin{equation}
\frac{\partial V_\text{DGB}}{\partial \phi_j} =\sum_i \frac{\partial V_\text{DGB}}{\partial \mathcal{I}_i} \frac{\partial \mathcal{I}_i}{\partial \phi_j}=\sum_j \frac{\partial V_\text{DGB}}{\partial \mathcal{I}_i} J_{i j}=0\,, 
\end{equation}     
where $J_{ji} = \partial \mathcal{I}_i/\partial \phi_j$. Extrema that happen when $\partial V_\text{DGB}/\partial \mathcal{I}_i = 0$ appear in specific realizations of the theory and are functions of the $c_{abc}$'s. We are interested in those extrema which correspond to taking $J_{ji} = 0$, and label them as \emph{maximally natural extrema} (MaNa extrema), since they only depend on the underlying symmetry of the theory.

Fig.\,\ref{fig:invariant_manifold} shows the manifold defined by the invariants $\mathcal{I}_6$ and $\mathcal{I}_{10}$ of $A_5$ as they take all their possible values. The MaNa extrema are located on the 0-dimensional intersections between the different 1-dimensional branches of the boundary of the manifold, where the jacobian must always vanish.
It can be shown that around each of those extrema, a smaller symmetry remains explicitly realized. In the next subsection, we briefly review the phenomenological consequences of the $Z_5$ invariance of point $B$.

\paragraph{Phenomenology}

\begin{figure}
\centering
\begin{subfigure}{0.4\textwidth}
\includegraphics[width = \textwidth]{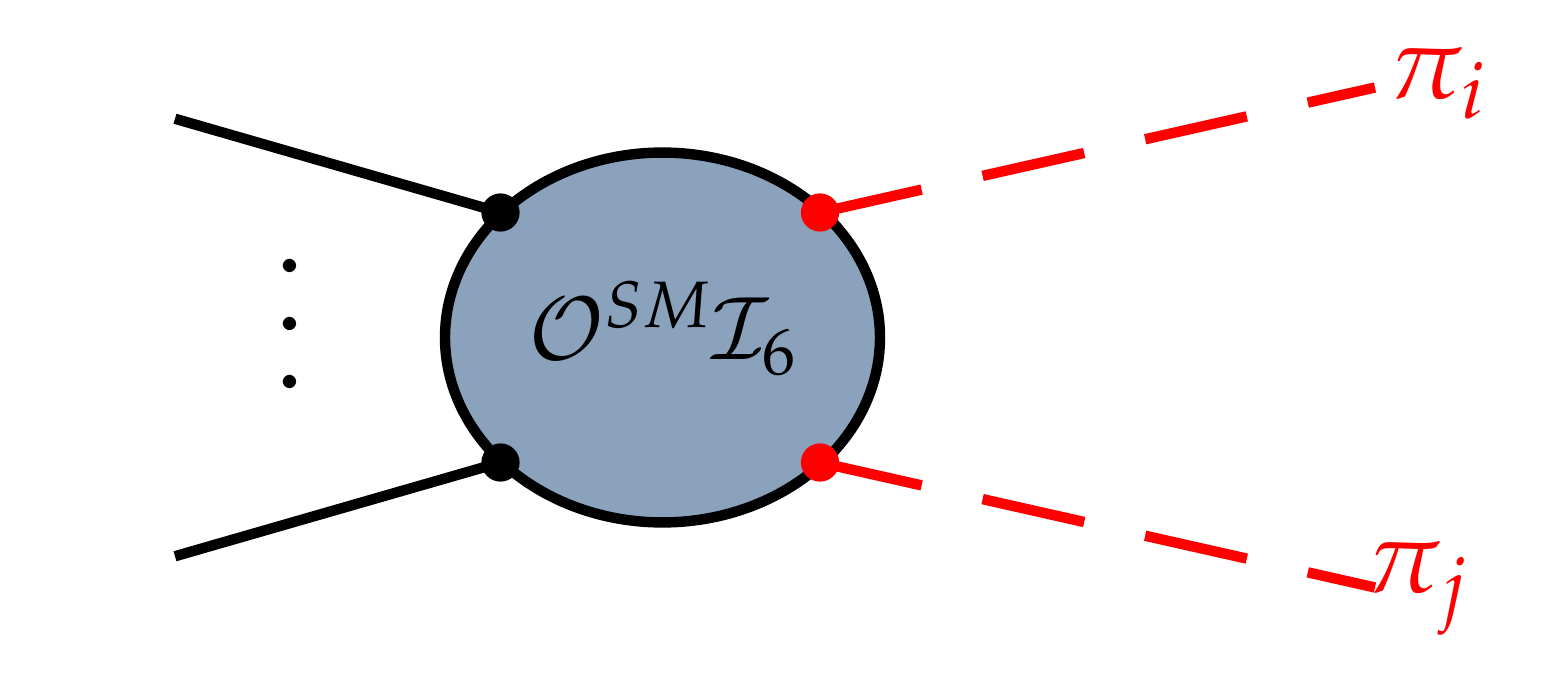}
\end{subfigure}
\begin{subfigure}{0.4\textwidth}
\includegraphics[width = \textwidth]{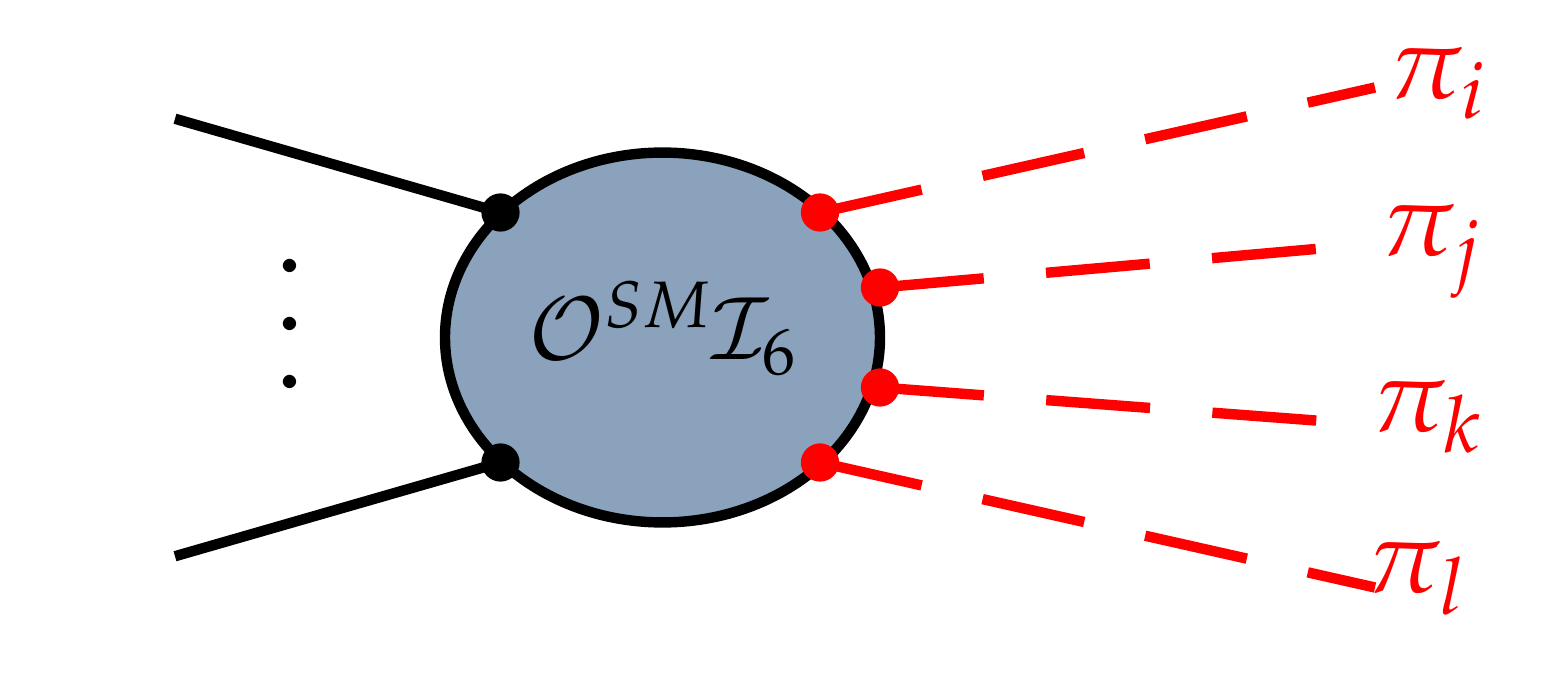}
\end{subfigure}
\begin{subfigure}{0.4\textwidth}
\includegraphics[width = \textwidth]{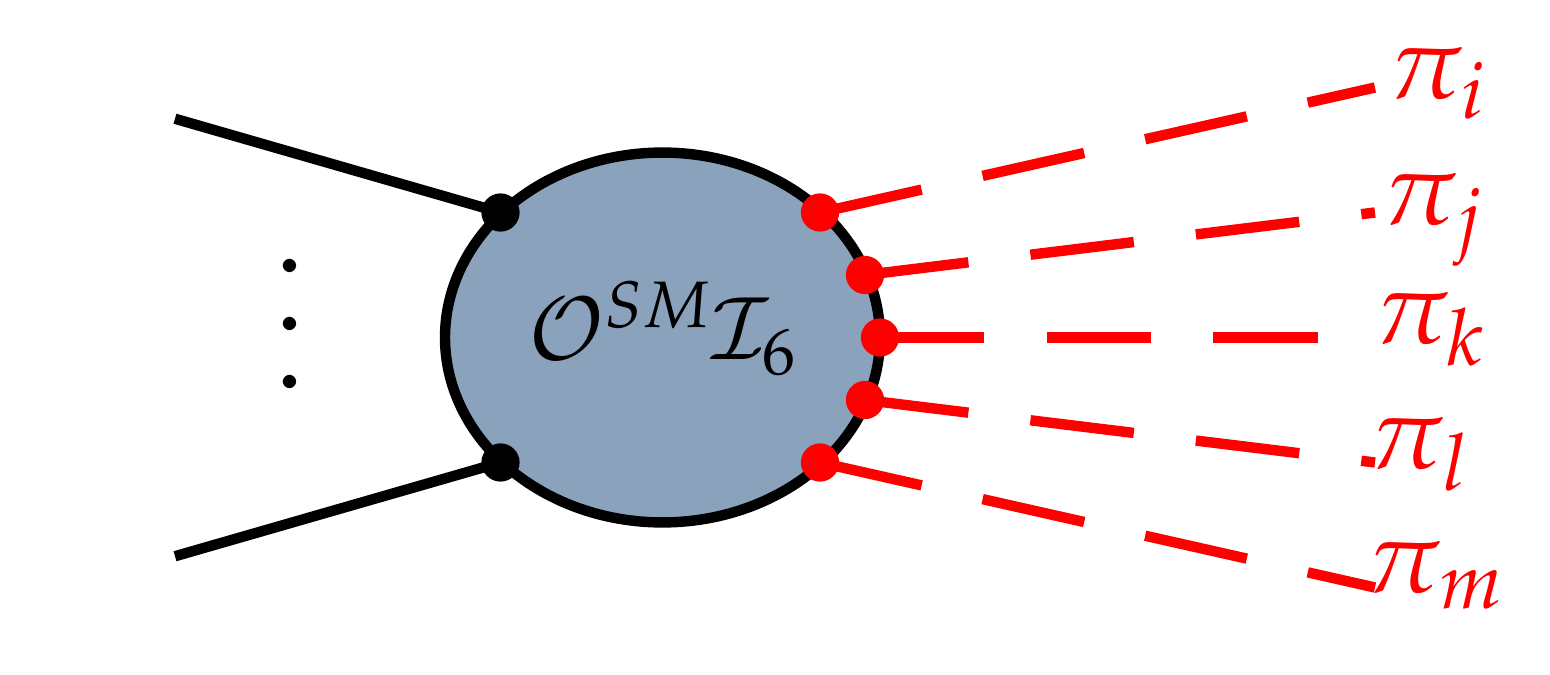}
\end{subfigure}
\caption{Production of DGBs from SM collisions, for a triplet of non-linearly realized $A_5$. Notice that events with three DGBs are absent.}
\label{fig:invariant_manifold}
\end{figure}

The dominant contribution in Eq.\,(\ref{eq:A_5:invariants}) is simply given by the invariant $\mathcal{I}_6$; expanding it around point $B$ in Fig.\,\ref{fig:invariant_manifold} after SSB we find
\begin{equation}
\mathcal{I}_6 =\frac{32}{5} f^4\left[\frac{f^2}{32}+\left(\pi_1^2+\pi_2^2\right)-\frac{31}{12 f^2}\left(\pi_1^2+\pi_2^2\right)^2-\frac{1}{4 f^3}\left(\pi_1^5-10\, \pi_1^3 \pi_2^2+5\, \pi_1 \pi_2^4\right)\right]\,.
\label{eq:A_5:I_6_on_B}
\end{equation}
The DGBs, $\pi_1$ and $\pi_2$, appear arranged in combinations, $\mathcal{J}^{Z_5}_2 = \pi_1^2 + \pi_2^2$ and $\mathcal{J}^{Z_5}_5 = \pi_1^5-10\, \pi_1^3 \pi_2^2+5\, \pi_1 \pi_2^4$, which are invariant
under $Z_5$. This leads to distinctive phenomenological consequences. Firstly, the two DGBs are degenerate, as it can be read out from the mass term in Eq.\,(\ref{eq:A_5:I_6_on_B}). More interestingly, $Z_5$-invariance also determines which are the interaction terms in the theory and their relative weights. This leads to definite ratios among production rates $\sigma(SM\rightarrow n\pi)$ of different numbers of DGBs from the SM. Under the assumption that the SM is a singlet of $A_5$, we have the ratios
\begin{equation}
    \frac{\sigma(\mathrm{SM} \rightarrow 2 \pi)}{\sigma(\mathrm{SM} \rightarrow 4 \pi)}=\frac{216(4 \pi)^4}{19(31)^2} \frac{f^4}{E_{\mathrm{CM}}^4}\,, \quad \frac{\sigma(\mathrm{SM} \rightarrow 4 \pi)}{\sigma(\mathrm{SM} \rightarrow 5 \pi)}=\frac{19(31)^2(8 \pi)^2}{(45)^2} \frac{f^2}{E_{\mathrm{CM}}^2}\,,
\end{equation}
which may serve to disentangle $A_5$-symmetric DGBs from different realizations of DGBs or other BSM scalar particles. Notice as well that, since the exact $Z_5$ forbids terms linear in the DGBs from appearing the Lagrangian, any process involving these particles must have them in pairs at least. This is in contrast with e.g. axions or ALPs, and provides yet another handle to differentiate DGBs from other scalar particles.

\section{Conclusions}

We have summarized the work in \cite{Vileta:2022jou} on discrete Goldstone bosons. The main theoretical motivation of this novel class of particles is the forbidding of quadratic corrections with UV scales, which is granted automatically upon the preservation of an exact discrete symmetry. Interesting phenomenological consequences ensue, which are generic for all DGBs and we have presented for the compelling case of a triplet of $A_5$.

\bibliography{bibliography.bib}
\bibliographystyle{utphys.bst}

\end{document}